\documentclass[12pt]{elsart}
\usepackage{graphicx}
\usepackage{amssymb}
\usepackage{color} 
\begin{document}
\begin{frontmatter}
\title{Simulation studies of CZT Detectors as Gamma-Ray Calorimeter}

\author[a1]{I. Jung \corauthref{cor1}}
\ead{jung@physics.wustl.edu}
\corauth[cor1]{I. Jung}
\author[a1]{H.~Krawczynski}
\author[a1]{S.~Komarov}
\author[a1]{L.~Sobotka}

 \address[a1]{Washington University in St. Louis
   Department of Physics
   1 Brookings Dr., CB 1105
   St Louis
   MO 63130}

\begin{abstract}
Several astrophysics and nuclear physics applications require the
detection of photons in the energy range of keV up to several MeV with 
good position and energy 
resolution. For certain applications Cadmium Zinc Telluride (CZT) detectors
might be the detector option of choice. Up to now, CZT detectors 
have mainly been used in the energy range between a few keV and $\sim$1~MeV.
They operate at room temperature and
achieve excellent  position resolution and substantially better 
energy resolution than 
scintillation detectors. Furthermore, they can be built more compact and
more economically than Ge detectors and do not require cryogenic cooling.

In this paper, we describe the results of 3-D Monte Carlo simulations
of a ``CZT calorimeter'' that can be used to detect  photons in 
the keV to several MeV range.
 The main objective of these studies is to evaluate the feasibility of CZT 
calorimeters, 
to study their performance and detect and understand 
performance limiting factors. Such a calorimeter consists of many layers
of closely packed pixellated CZT detector units. 

Our simulations of single detector units reproduce experimental  
results, indicating that our simulations capture the main factors that
limit the performance of a detector unit. 

For a full calorimeter the limiting factors within a range from $\sim$20~keV to $\sim$10~MeV are: a) the fact, that the incident energy is not totally deposited  
within the detector area because secondary particles leave the detector against the direction 
from which the incident radiation enters, b) signal loss when the interaction is near the pixel edges and near the anodes. In this case signals which are 
induced in neighboring pixels are discarded when their intensities lie below the trigger threshold. c) 
the steep weighting potential gradient close 
to the anodes, which affects about 0.25~cm next to the anode and impairs there 
 the correction of the depth of interaction (DOI). This effect dominates in thin detectors (0.5~cm).\\

Understanding the limiting factors we come to the conclusion that
1~cm to 1.5~cm thick 
detector units can be used to build a calorimeter with good performance over 
the energy range from $\sim$20~keV to $\sim$10~MeV .

\end{abstract}

\end{frontmatter}

\section{Introduction}
\label{Introduction}
Cadmium Zinc Telluride (CZT) has emerged as the detector
material of choice for the detection of hard X-rays and soft $\gamma$-rays
when excellent position and energy resolution is needed and 
cryogenic cooling is impractical.
Most commonly, CZT detectors are employed to detect
photons in the 10 keV to $\sim$1 MeV energy range where its high density
($\rho\,\approx$ 5.76 g/cm$^3$)  
and high average atomic number ($\simeq$50) result in
high stopping power and a large cross section for photoelectric
interactions.
Several astrophysical, nuclear physics, and homeland security applications
require the detection of photons with energies of several MeV 
with good position and energy resolution and with a detection efficiency
close to 100\%.  A detector built from closely packed CZT detectors
units may perform
substantially better than a scintillation detector and may be more compact
and more economic than a cryogenically cooled Ge detector.
So far 
CZT detectors have mainly been studied in the photoelectric regime
(10 keV-$\sim$300 keV). 
Two different designs have been widely used, pixellated  detectors and 
single sided strip detectors \cite{He2005,Matt,Macri,DOE}. 
Excellent energy resolutions have been achieved with both designs.
 For a pixellated detector 
with 11 $\times$ 11 pixels and a steering grid Zhang et al. \cite{He2005} reported 0.8\% 
(full width half maximum, FWHM) at 662 keV for single pixel
events and 2.3\% at 662 keV (FWHM) for three pixel events. The detector size was 1.5  $\times$ 1.5 $\times$ 1 cm$^3$.
For a single sided charge-sharing CZT strip detectors FWHM energy resolutions
of 19.5 keV at 662 keV  and 23.7 keV ar 1333 keV were reported \cite{DOE}. 

Whereas detectors with N $\times$ N pixels require N$^2$ readout 
channels CZT cross-strip detectors require 2 $\times$ N channels. 
The penalty for this advantage are ambiguities in matching strips with the 
right cross-strip counterparts
in case of multiple
interactions in one detector.
In order to achieve good detector efficiency, energy resolution 
and imaging properties in the  ``Compton'' regime of several hundred keV 
requires
to take multiple pixel or multiple strip events into account. Strategies exist to ameliorate this problem somewhat, for further details \cite{DOE}.
This paper deals with pixellated 
CZT detectors only. We use  the Geant 4.0 code \cite{G4}  to study the
performance of very thick detectors ($\gg$ 10 cm) built from closely packed
CZT detector units with the objective of evaluating the
theoretically achievable performance of such CZT calorimeters given
the electronic properties of present-day CZT substrates. As we will
describe further below, the main thrust of our study is to evaluate
general performance limitations arising from the combination of  
the location and spatial extent of the charge clouds below individual 
detector pixels, trapping of electrons and holes and the influence of the 
weighting potential.
We neglect the performance
limitations arising from Te-precipitates and other crystal defects. 
The good agreement of the simulated detector response and the one 
experimentally measured shows that electron and hole trapping and the
detectors weighting potential indeed limits the performance of current
CZT detectors. Te-precipitates and other crystal defects seem to be only
important for underperforming detector areas. 

The calorimeter performance depends on the shape of the 3-D weighting potential and on the number,  energy and distribution of secondary particles 
produced in the calorimeter. Simulations with a particle interaction
code like Geant 4 and a device simulation code is the most efficient and most accurate way to estimate the performance of such calorimeters.

This paper is structured as follows. In Section
\ref{AstroNuclAppl}, we give a brief
description of astrophysics and nuclear physics experiments that may
use CZT calorimeters. 
Subsequently, in Section  \ref{CZTSimulation} we describe the simulations 
and present their results for single detector units and for 
full CZT calorimeters.
In Sect. \ref{con} the results are discussed. 

In the following all energy resolutions are given as Full-Width-Half-Maximum
energy resolutions.
 
\section{Applications in Astrophysics and Nuclear Physics}
\label{AstroNuclAppl}
\subsection{Astrophysics}
From several hundred keV to $\sim$10~MeV the dominant interaction process of
photons in matter is Compton scattering. At lower  energies the
photo effect dominates and at higher energies pair\- production. The sensitivity of past, present and upcoming 
hard X-ray- 
and $\gamma-$telescopes shows a gap in the ``Compton'' energy range 
from a few 100 keV up to several tens of MeV. 
This energy range is highly interesting from the astrophysical point of view.
Some of the scientific objectives are: the study of nucleosynthesis in
core-collapse, type Ia supernovae and classical novae, 
 the search for ``intermediate`` blazars with peak emission of 
about 1 MeV and $\gamma-$ray lines from nuclear interactions of 
cosmic rays with the interstellar medium \cite{MEGAProject}, and the 
study of the 511 keV emission from sources like the galactic center
\cite{Knoe}. \\[2ex]
\begin{figure}
\centering
\includegraphics[width=2.5in]{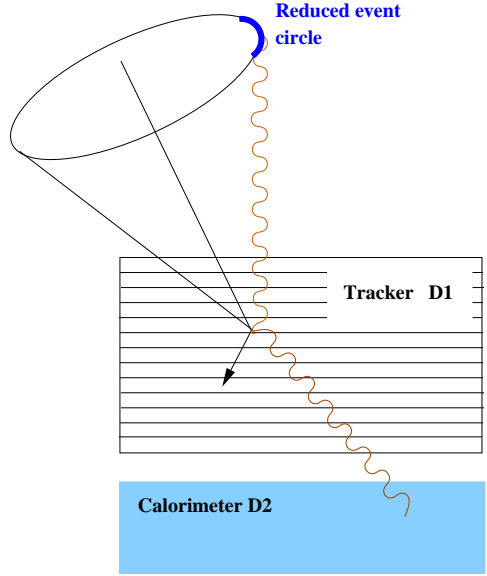}
\caption{Sketch of a Compton $\gamma$-ray telescope operating in the 
energy range from a few hundred keV to several ten MeV. 
Measuring the position and energy deposition in D1 and in D2, the incident 
direction of the primary $\gamma-$ray can be constrained to lie
on a circle. Additional measurements of the direction of the 
recoil electron would make it possible to constrain the incident direction
to an arc. }
\label{DesigneConcept}
\end{figure}

Figure \ref{DesigneConcept} shows the basic detector principle used in both the
TIGRE (tracking and imaging  $\gamma-$ray instrument) \cite{TIGERProject} 
and the MEGA (medium energy $\gamma-$ray astronomy) designs 
\cite{MEGAProject}. The design uses two 
detectors: first, a tracker (D1) where Compton scattering and pair conversion 
takes place and where the trajectories, energies, and momenta of 
secondary electrons are measured; second, a calorimeter (D2) 
to determine the location and energy of scattered photons 
(see Figure \ref{DesigneConcept}).
The Compton equation can be used to calculate the energy and incident 
direction of the primary photon from the combined information from the D1 
and D2 detectors. Depending on the knowledge of the momentum and energy 
of the scattered electron, the direction of the incident photon is given 
either as an arc or a circle. Usually a low-Z material is used for the D1 
detector to maximize the probability of a single Compton scattering. 
For D2
a high-Z material is the preferred choice to maximize 
the probability for photon absorption in photo effect events.

In the specific case of MEGA, the tracker 
consists of 32 layers of double-sided silicon strip detectors 
(6 $\times$ 6 cm$^2$ each, 500 $\mu$m thick, with
a pitch of 470~$\mu$m). The calorimeter consists of CsI slabs, 
5 $\times$ 5  cm$^2$ cross section, 4-8~cm deep. 
Each slab is read out by a Si PIN diode.
Over the energy range from 511 keV to 1.274 MeV, 
the CsI calorimeter 
achieves an energy resolution of about $\sim$8\% FWHM and
 a spatial resolution of 2 cm \cite{MEGACalib,MEGACalorimeter}.
\\
\\
Ge calorimeters could in principle be used to improve the performance  of Compton telescopes like TIGRE and MEGA, however, the costs of constructing
such a calorimeter and  the cryogenic cooling
system, and lifting it into an orbit are very high. 
 CZT with excellent spatial resolution and good energy resolution
 might be a better choice for D2 to improve on
the performance of Compton telescopes.

\subsection{Nuclear Physics}
Another application for CZT calorimeters lies in the detection of MeV 
photons in nuclear physics experiments that use ``fast beams'' 
($\beta = v/c \,>$ 0.2). Presently, Ge detectors are used to 
measure the energy and location of $\gamma-$rays emitted by excited nuclei. 
However, Ge detectors are  expensive
and  require cryogenic cooling that makes them bulky, and complicates the use 
with auxiliary detectors or spectrometers.
In the case of ``fast beams'', the  effective energy resolution of the experiment 
(the accuracy of the energy reconstruction in the rest frame of the emitting 
particle) is mainly limited by the spatial 
resolution with which the $\gamma-$ray can be localized, owing to the
Doppler effect. 

\begin{figure}
\centering
\includegraphics[width=3.0in]{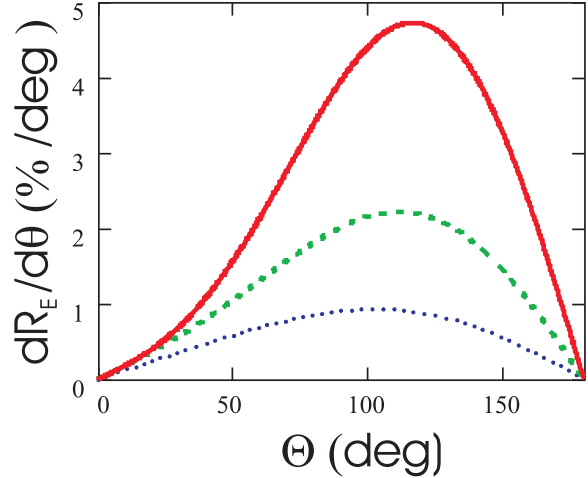}
\caption{Differential energy change  $dR_{E}/d\theta$ with angle $\Theta$
between emitter and photon momenta for beta = 0.25 (dotted line), 0.50 (dashed line) 
and 0.75 (solid line). If the spatial resolution of the detector allows one
to measure the arrival direction of the photon with an accuracy of $\Delta\theta$, the associated uncertainty in the energy of the photon in the momentum
of frame is $dR_{E}/d\theta\Delta\theta $.
}
\label{Doppler}
\end{figure}

Figure \ref{Doppler} shows the differential energy change $dR_{E}/d\theta$ 
as a function of
the angle between  emitter and photon momenta for $\beta$ = 0.25, 0.50 
and 0.75, dotted, dashed and solid line, respectively. For an angle 
of 50$^{\circ}$ the differential energy change is $\sim$~0.5\%/degree,
$\sim$~1\%/degree and $\sim$~1.4\%/degree for
 $\beta$ = 0.25, 0.50 and 0.75, respectively. 
Thus, in fast beam experiments the effective energy resolution of the experiment 
is limited by the spatial resolution of the
detectors. \\
The Ge detector GRETA (Gamma-Ray Energy Tracking Array) 
is the next generation $\gamma-$ray detector for 
nuclear structure research. 
GRETA will use a two-dimensionally segmented Ge detector along with 
pulse-shape analysis
to determine the three-dimensional position and the energy of
the interaction. The prototype GRETINA  of GRETA 
\cite{Gretina,Gretina1}
consists of a cluster of three tapered HP-Ge n-type crystals 
with a hexagonal
shape. The crystal length
is 9 cm, the diameter at the back is 8 cm. 
 The outer electrode is divided into 36 parts,
6 longitudinal and 6 transverse segments.
GRETINA achieves laboratory frame energy resolutions of 1.15 keV at 60 keV and 2.35 keV at 1.3332 MeV.  The position resolution 
is 2 mm RMS in three dimensions\footnote{$http$://$fsunuc.physics.fsu.edu/$$\sim$$gretina/Newsletter4/GRETINA\_newsletter\_4$}. 
For a CZT detector,
the theoretical limit of the 
position resolution is the distance electrons produced 
in photo effects or Compton events travel before transferring their energy
to the valence band electrons.
At 1 MeV 80\% of the electrons stay within 0.18~mm  from the point where they are generated. 
For real pixellated detectors the position 
resolution is limited further by pixel size, timing and possibly
cathode signal resolution. Zhang et. al \cite{He2005} reported position resolution
of 1.27 $\times$ 1.27 $\times$ 0.2 mm for a 1.5 $\times$ 1.5 $\times$ 1.0~cm$^3$ CZT detector with 11 $\times$ 11 pixels. Despite the clearly better energy resolution of a Ge calorimeter low cost small volume CZT calorimeters can therefore achieve comparable overall effective energy resolution in fast beam experiments.

\section{Simulation Study of a CZT Calorimeter}
\label{CZTSimulation}
In the recent years the performance of pixellated CZT detectors has
been dramatically improved. Energy resolution at 662~keV
of 0.8\% for single pixel readout and 2.3\% for combined signals 
of three anode pixels were reported for a CZT detector of dimension 
1.5 $\times$ 1.5 $\times$ 1.0~cm$^3$ with 11 $\times$ 11 pixels \cite{He2005}.
For the detection of higher energies it is necessary to 
combine the signals of several pixels because the dominant interaction process, the Compton effect, tends to spread the 
interaction over adjoining pixels. Even at lower 
energies integrating over all pixel signals has a large impact 
on detector efficiency, because the incidents of Compton events is not negligible.

\begin{figure}
\centering
\includegraphics[width=3.0in]{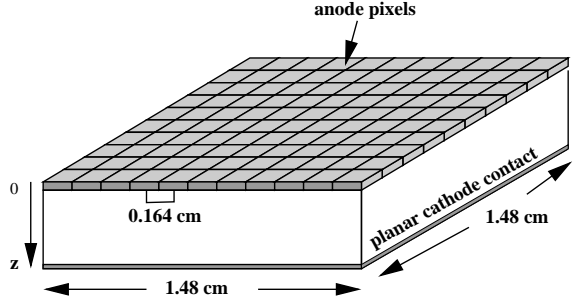}
\caption{The sketch shows the layout of the simulated single 
detector unit and  
the coordinate system used in the text. 
The thickness of the detector is 0.5 cm, 1.0 cm or 1.5 cm, 
the pixel pitch 0.164  cm and the pixel width 0.162 cm.}
\label{SingleDetector}
\end{figure}

To gain a better understanding of the behavior 
of pixellated CZT calorimeters, detailed three dimensional
simulations of a large volume calorimeter have been performed. 
Our main objective is to derive theoretical limits for currently 
used CZT detectors and electronics. Therefore we have chosen to 
use a CZT calorimeter constructed out of 1.48 $\times$ 1.48~cm$^2$ 
single pixellated CZT detector units  with 9 $\times$9~pixels and a pitch 
of 0.164~cm. Reported pixel pitches vary between 0.11~cm and 
0.25~cm.  The pixels are held at ground and the cathode bias
is -1500~V per 0.5~cm. The thickness of the detector units 
was varied: 0.5~cm, 1.0~cm and 1.5~cm. All these sizes are 
commercially available.
The overall size of the calorimeter considered in the following 
 was chosen to be  10.33~$\times$~10.33~$\times$~45~cm$^3$.
This choice ensures  
that almost all interactions take place inside the detector volume.
The results are not limited to this particular design but apply to 
to all calorimeters as long as they are large enough to fully stop incoming 
$\gamma-$rays. 

The output of our simulations are electron drift times and signals on all pixels. By ``signal'' we understand the integrated charge over 
1~$\mu$s. If not stated otherwise, we assume 
 electronic noise of 4.5~keV and a pixel trigger 
threshold of 13.5~keV. These values are typical for state-of-the-art readout systems \cite{He2005}. 
We did not simulate passive material (printed circuit boards, 
ASICs) between the CZT detector units. 

The simulations can be separated in three parts. 
First the three dimensional electric field and the weighting potential 
are calculated with  a finit-difference method.
The electric field is calculated at the nodes of a 3-dimensional grid with
the grid size $\Delta$s. If the coordinates of the nodes are given by 
(i$ \cdot \Delta$s, j$\cdot \Delta$s, k $ \cdot \Delta$s), the Laplace 
equation for each node with unknown potential can be 
approximated by the central point finite difference scheme:
\begin{eqnarray}
  \epsilon_{i+1,j,k}(\varphi_{i+1,j,k}-\varphi_{i,j,k})&-
  \epsilon_{i-1,j,k}(\varphi_{i,j,k}-\varphi_{i-1,j,k})&+\\\nonumber
  \epsilon_{i,j+1,k}(\varphi_{i,j+1,k}-\varphi_{i,j,k})&-
  \epsilon_{i,j-1,k}(\varphi_{i,j,k}-\varphi_{i,j-1,k})&+\\\nonumber
  \epsilon_{i,j,k+1}(\varphi_{i,j,k+1}-\varphi_{i,j,k})&-
  \epsilon_{i,j,k-1}(\varphi_{i,j,k}-\varphi_{i,j,k-1}) &=0
\end{eqnarray}
After specifying the ``boundary conditions'', by setting the potential 
of the contact nodes to the applied bias potential, one gets a system
of linear equations of the form
\begin{equation}
   \sum_{l=1}^{N}a_{i,l}\cdot\varphi_{l} = b_{i}
\end{equation}
where the index l runs over the remaining N nodes where the potential
has yet to be determined.
We use the
steepest descent method, to solve this system of linear equations numerically. 
This iterative method minimizes the difference $\sum_{l=1}^{N}a_{i,l}\cdot\varphi_{l} - b_{i}$
until it becomes smaller than a certain preset value. 
To determine the weighting potential of pixel i, the potential of the contact
node of i is set to 1, for the rest to 0.
\\
Second, for the interaction of incident $\gamma-$rays and secondary particles 
with the detector material the  low-energy extension of 
the Geant 4.0 code \cite{G4} is used. For each interaction the electrons
and holes are  
tracked in 1~ns steps through the detector. For each step, the 
induced charge at the anodes is calculated. The rise time of each pixel
and the charge accumulated in 1~$\mu$s are the resulting signal values.
The electron mobility is set to
$\mu_{e} = 1000$~cm$^2$/V/s  with a lifetime of $ \tau_{e} = 3\:\mu$s and the
hole mobility to $\mu_{h} = 65 $cm$^2$/V/s with a lifetime of 
$ \tau_{h} = 1\:\mu$s \cite{evProduct}.
The simulations neglect any effect of material non uniformity.

\subsection{Results for a single detector unit}
For the understanding of a complex system it is often useful to study first  a small part instead of the full system. Thus
we begin with the discussion of a single CZT detector unit.

\begin{figure}
\centering
\includegraphics[width=5.5in]{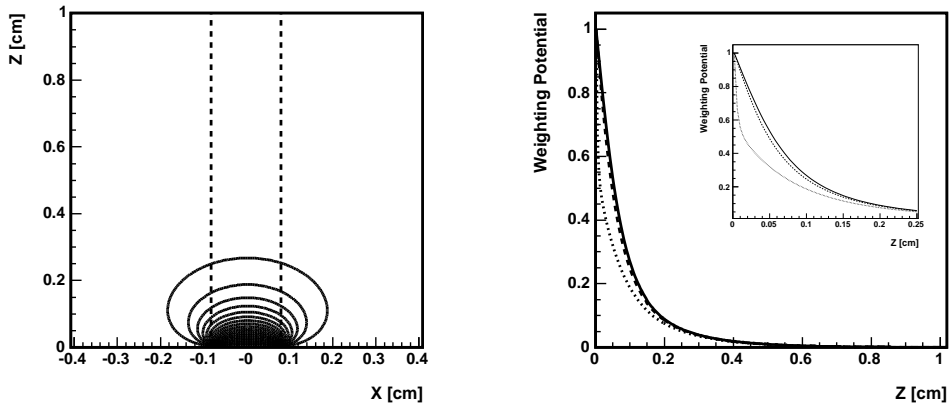}
\caption{On the left side is the weighting  potential distribution
of the central pixel. Shown is a cross section along the xz-plane through
y=0. The pixel extends from -0.082~cm to 0.082~cm, 
marked by the two dashed lines. On the right side 
the weighting potentials distribution as a function of z is given 
for three lateral positions (x=0~cm (solid line), x=0.04~cm 
(dashed line) and x=0.08~cm (dotted line), y=0~cm). The overlayed histogram  zooms in
to z-values between 0~cm and 0.25~cm.}
\label{WP}
\end{figure}

Figure \ref{WP} left panel shows the potential distribution 
of the central pixel for a 1~cm thick 
detector.  The weighting potential distribution for three 
different lateral positions as a function
of z (equal depth of interaction  (DOI) ) is given in Figure 
\ref{WP} right panel. 
One can see, that most of the signals are
induced near the anode contacts (so called small pixel effect) \cite{Scho,Ramo}, 
and the anode signal depends only weakly on the depth of interaction (DOI),
 if the interaction takes place less than 0.6~cm away from the 
cathode. The residual DOI dependence is caused by small changes 
of the weighting potential far away from the pixels and by
electrons getting trapped as they drift to the anode. 
Near the anode the weighting potential rises steeply.
This causes a strong dependence of the induced charge in this 
region. Interactions  closer to the anode induce less charge than 
interactions closer to the cathode. It can be also seen, that the 
dependence in x-y direction increases near the anodes too. 
The cathode behaves differently since it does not exhibit the small
pixel effect.

CZT is a single charge carrier, therefore it is necessary to correct for the 
DOI. For single-pixel events, one can derive the 
depth of interaction by using 
the cathode-to-anode signal ratio. This approach does not work 
for multi-pixel events. In this case,  the signal rise time 
(proportional to the electron drift time) can be used \cite{He2005}. 
This is the approach we used in our simulations. 

\begin{figure}
\centering
\includegraphics[width=2.5in]{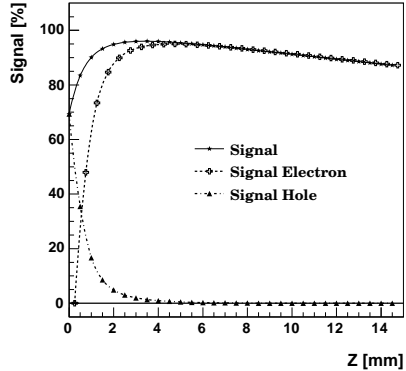}
\caption{The percentage of energy deposition 
as a function of depth of interaction is shown for a single detector, 
1.5~cm thick. 
Energy deposition in the pixel center was simulated for the central pixel. 
The results are presented for a cross section along the xz plane through y = 0.}
\label{ZDep}
\end{figure}

Figure \ref{ZDep}, shows the anode signal as
the percentage of the energy  of the incident gamma as a function of DOI.
The behavior can easily be understood based on the charge transport 
properties of CZT and the weighting potential (Figure \ref{WP}). 
The gradient of the weighting  potential is very steep close to the 
anodes and flattens with increasing z. For interactions
close to the anodes, electrons drift only small distances before 
they impinge on the anodes. Holes contribute up to 70\% of the total 
signal by moving through the steep gradient of the weighting 
potential. For interactions closer to the cathode side, electrons induce
charge as they move towards the pixels. Holes induce relatively little 
charge, as they are trapped within $1\mu$s (0.12~cm) of being 
generated, and move towards the cathode through a region of rather
constant weighting potential. 

The relative signal contribution of electrons and holes depends on 
the charge integration window. If the integration time is shorter
the holes travel less distance and thus their contribution
is smaller.

\begin{figure}
\centering
\includegraphics[width=6.0in]{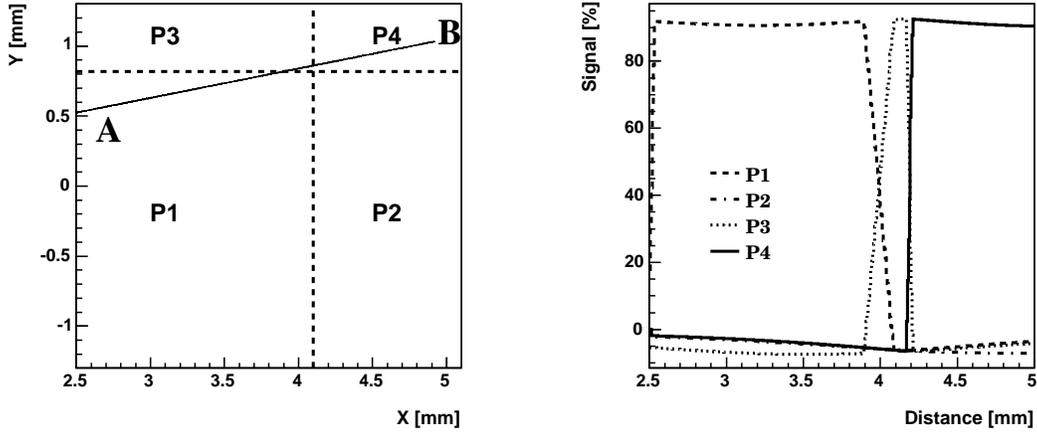}
\caption{On the left side, the inner four pixels of the detector (P1, P2,
P3, P4) are shown with a line where charge deposition was simulated.
The right side shows the charge induced on the four pixels (in units of the 
deposited charge) for different deposition locations along the line
from point A to B.}
\label{InducedSignal}
\end{figure}

The weighting potential is not confined to the volume of the primary affected pixel. (see 
Figure \ref{WP}) Therefore neighboring pixels measure signals, when an
interaction takes place. The amount of induced charge depends 
on the point of interaction. The closer it occurs to the pixel edges and to the anodes the larger is the signal in the 
neighboring pixel. To explore this effect in more detail, Figure \ref{InducedSignal}, right panel, presents the 
signals induced in four adjacent pixels as a function of the point of interaction which is indicated as a line on the left panel. 
One can nicely see how the signal is spread over adjacent pixels
when charge is deposited close to the pixel boundaries.

\begin{figure}
\centering
\includegraphics[width=6.0in]{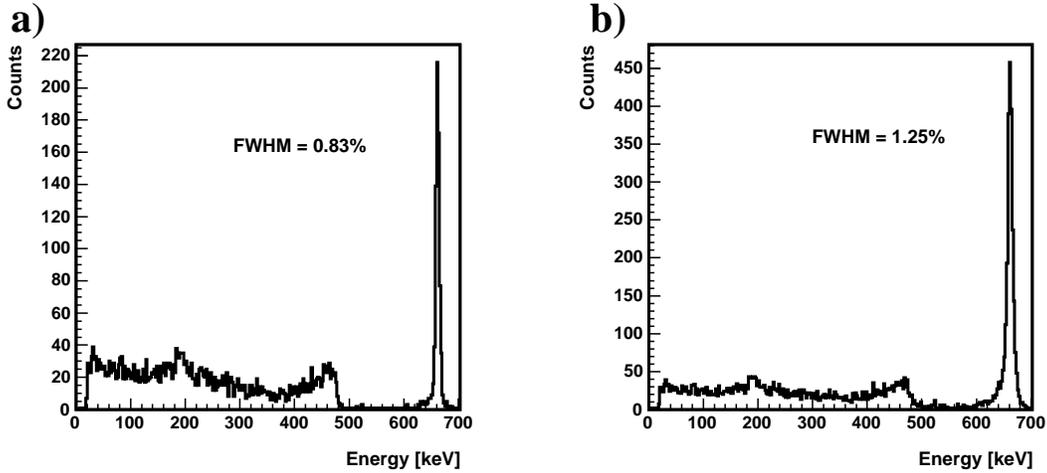}
\caption{Simulated 662~keV energy spectra of a single detector unit.
The left panel shows single-pixel events. The right panel shows the 
energy spectra for all events registered by the detector. The method used
for reconstructing the energy of the primary 
$\gamma-$ray is described in the text. All $\gamma-$rays were assumed to hit 
the detector's central pixel. 
}
\label{1cmDetectorSpectra}
\end{figure}

Current experiments with pixellated CZT detector report energy 
resolution of 0.8\% for single pixel \cite{He2005} and
2.3\%  for the combined signals of three pixels at 662~keV. The detector
size was 1.5 $\times$ 1.5 $\times$ 1~cm$^3$ with 11 $\times$ 11 pixels
and a steering grid. The electronic noise was in the range of 
4.5~keV. For comparison we have simulated
the energy resolution of a single detector unit, 1~cm thick, 
both for single-pixel events and for combining all anode signals. 
DOI correction is applied. 
The spectra obtained are given for 662~keV in Figure 
\ref{1cmDetectorSpectra}.

 The energy resolution of single-pixel events
is 0.83\% while for all events including multiple-pixel events,
 it is 1.25\%. 
For single-pixel events the peak to Compton ratio is about $\sim$20
and for the full detector $\sim$25. These results are in good agreement
with the data  reported by Zhang et al.
This comparison shows us, that the detector performance is 
determined by CZT general properties and detector design and not by the incidental quality of the CZT material of a specific detector.
\\

Currently commercially available detectors are 0.5~cm, 1.0~cm 
or 1.5~cm thick. The results of our simulations  for 662~keV incident energy are given in Table 
\ref{tab:energy}.
\begin{table}[htbp]
  \centering
  \begin{tabular}[h]{|l|l|l|}\hline
    thickness $[$cm$]$ & FWHM $[$\%$]$ & FWHM  $[$\%$]$\\ 
    & single-pixel events & multi-pixel events \\ \hline
    0.5 & 1.12 & 1.34 \\\hline
    1.0 & 0.83 & 1.25 \\\hline
    1.5 & 0.80 & 1.25 \\\hline
  \end{tabular}
  \caption{Energy resolution for 662~keV incident 
    energy for single-pixel and 
    multi-pixel events for three detector thicknesses of a single CZT detector unit.} 
\label{tab:energy}
\end{table}

For 0.5~cm and 1.5~cm thick  detectors we obtained single-pixel energy 
resolutions of
1.12\% and 0.80\%, 
respectively and multiple pixel 
energy resolutions of 
1.34\%  and 1.25\%.
While 1.0~cm and 1.5~cm thick CZT detector units show almost identical behavior, the energy resolution of a 0.5~cm thick detector drops to 1.12\%.

\begin{figure}
\centering
\includegraphics[width=6.0in]{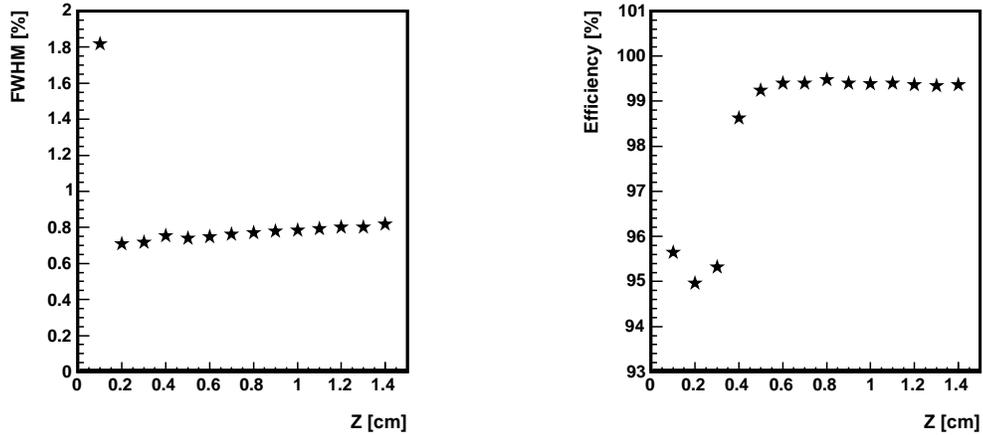}
\caption{On the left side, the energy resolution 
of a central pixel is shown as a function of 
 distance from the anode side z
for 662~keV incident energy and the 1.5~cm thick detector. 
On the right hand side shows the percentage of events within $\pm 3\sigma$
of the peak of the energy spectra.}
\label{AufloesungVSZ}
\end{figure}

In the following we will discuss how 
the energy resolution depends on a number of different effects. \\
We begin with an effect which results from the application of 
a trigger threshold.
When all DOI-corrected signals of all triggered pixels 
are summed up to calculate the total energy deposition,
signals below the trigger threshold will be discarded, 
diminishing the summed signal. This threshold effect becomes more 
prominent the closer the energy is deposited  to the anode. 
Therefore, the energy resolution depends on the DOI and thus on the 
 z-coordinate.
Furthermore, there exists a strong xy-dependence too, caused by the steep x- and y-components of the weighting potential gradient 
near the anode pixels. 
To study these two effects, we simulated charge deposition at specified z 
positions and random (x,y)-positions under one pixel. After applying the
 DOI correction we added all signals 
which exceed the trigger threshold (three times electronic noise). 
The results at 662~keV are presented in Figure 
\ref{AufloesungVSZ} for the 1.5~cm thick detector. 
On the left side, the energy resolution 
of a central pixel is shown as a function of the depth of 
interaction. 
On the right side, the percentage of events within the photo peak 
($\pm 3\sigma$) is shown. 
The energy resolution stays at 
about 0.8\% at distances between 0.2 to 1.5~cm from the pixel side. 
Closer to the anode, the resolution deteriorates to $\sim$1.8\%.
The fraction of events within the photo peak is 
$\sim$100\% for large z values
whereas it drops to about 95\% for events close to the pixel anode.

Energy spectra of events close to the anodes have a low energy tail.
Depending on the exact location below one pixel, a variable amount
of charge is induced on adjacent pixels. The charge is lost, if 
the adjacent pixel signal does not surpass the trigger threshold.

The results for 0.5~cm and 1.0~cm thick detectors 
are virtually the same as the corresponding region of the 1.5~cm thick 
detector, if one uses charge integration windows of the 
same duration. 

From Figure \ref{AufloesungVSZ} we see that the 0.25~cm region just below
the anodes gives very poor energy resolution. This for 0.5~cm thick detectors,
50\% of the detector volume performs very poorly. For 
1~cm and 1.5~cm thick detectors, the fraction of
poorly performing volume decreases to 25\% and 16\%, respectively.

\begin{figure}
\centering
\includegraphics[width=3.0in]{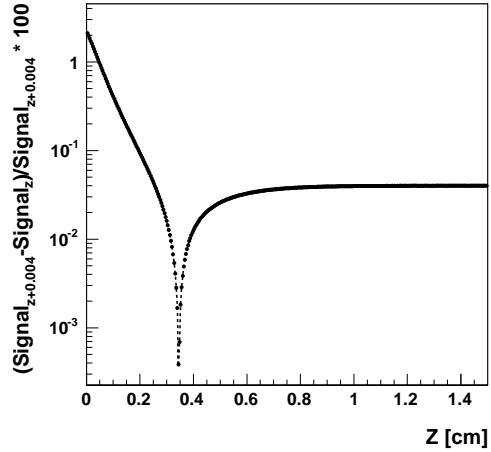}
\caption{The change in the measured signal over a 0.004~cm 
deep region along the z-axis is shown as a function of z. (x,y =0).}
\label{InteractionRange}

\end{figure}
The energy resolution is also limited by the fact that an incident
$\gamma-$ray can generate multiple energy depositions below one single 
pixel. As multiple charge depositions under one pixel  cannot 
be separated, the depth of interaction correction 
corresponds to the interaction with the longest drift 
time (closest to the cathode). However this correction is 
 applied to the full signal. At 662~keV the median  
 of the RMS values of the interaction region below one pixel with a 
thickness of 1.5~cm is 0.004~cm in z-direction.
We calculated the relative difference between the signals induced by 
a interactions at point z and at point z+0.004 with x- and y-coordinates set to 0. The result
is shown in  Figure \ref{InteractionRange} as a function of z, 
which gives the error in the DOI correction if one corrects 
for a DOI of z instead of z+0.004~cm.
Close to the pixel anode the charge spreading effect becomes more important 
because of the strong gradient of the weighting potential. 
The effect contributes considerably to the energy resolution below $<0.3$~cm.
At higher energies the  spreading effect is more pronounced and its influence on the energy resolution increases. 

To summerize the results above, the deterioration of the energy 
resolution for thinner CZT detector units is due to a) the steep 
gradient of the weighting potential close to the anodes, b) the induced 
signals in neighboring pixels below the trigger threshold and c) 
multiple interactions below one pixel.

\subsection{Results for a full calorimeter}
In this section we study the performance of a full calorimeter 
built out of several single detector units. The overall 
calorimeter 
size is $10.33 \times 10.33 \times 45 $~cm$^3$.
This calorimeter is made of several layers of CZT detectors. 
Each layer contains of an array of 
$7 \times 7$ detector units, each with $9 \times 9$ pixels.
As before, simulations were made varying the 
detector thickness (0.5, 1.0 and 1.5~cm). 
The overall size of the calorimeter was kept constant, 
therefore the number of layers was changed depending on thickness (90, 45 and 30 for 0.5~cm, 1.0~cm and
1.5~cm).

As before, we determine energy spectra by applying a DOI correction to 
each individual pixel signal satisfying the trigger
criteria. Afterwards all signals are summed up to obtain
the reconstructed energy.
 \begin{figure}
\centering
\includegraphics[width=5.5in]{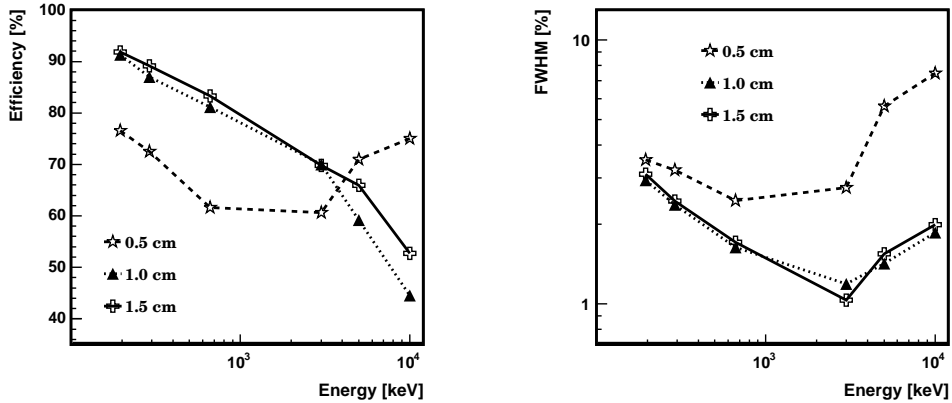}
\caption{On the left 
the detector efficiency is shown, i.e. the fraction of events 
 falling into the range of $\pm 3 \sigma$ from the  peak of the energy 
spectrum.
The right side shows the energy resolutions of  full calorimeters
 as function of incident energy for all three detectors designs.}
\label{EffizienzFWHM}
\end{figure}

The energy resolution is limited by several effects. Those which 
are due to the properties of the single detector units have been discussed 
in the previous section in some detail. Therefore we will limit the 
discussion here to effects which are peculiar to the full calorimeter.

The energy resolution of the full calorimeter as a function 
of incident energy are given on the right side of Figure  \ref{EffizienzFWHM}
for detector designs with base units of 0.5~cm, 1.0~cm and 1.5~cm 
thickness. 

\begin{figure}
\centering
\includegraphics[width=6.0in]{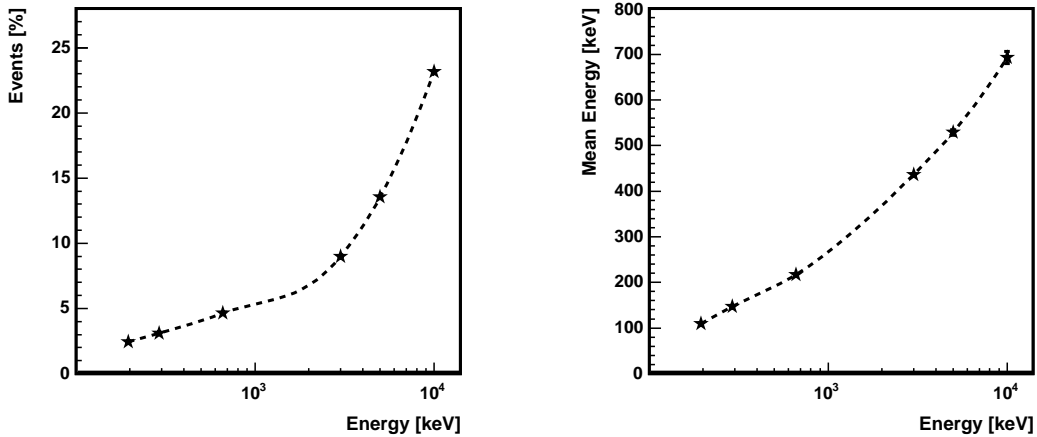}
\caption{The left side shows the percentage of events 
where secondary particles leave the detector in the upward direction 
as a function of incident energy. 
The right side shows the mean energy loss per event
with secondary particles leaving the detector in the upward direction
as a function of incident energy.}
\label{PercOutSide}
\end{figure}
On the left side, 
the detector efficiency is given, which we define as 
the fraction of events being reconstructed within $\pm 3\: \sigma$ 
of the peak of the energy spectra. Below 1~MeV a calorimeter 
made out of 0.5~cm thick detector units shows resolutions between 2\% and 5\%.
Above 1~MeV, the value steadily increases to $\sim$7.5\% at 10~MeV. The efficiency lies between 76\% and 60\%.
Calorimeters made of 1~cm and 1.5~cm thick detector units perform markedly
better. The energy resolution of both calorimeters improve 
from $\sim$3\% at 200~keV to $\sim$1\% at 3~MeV. At higher energies, 
the resolutions increase somewhat up to  $\sim$1.9\% at 10~MeV. 
The efficiency drops from $\sim$94\% at $\sim$200~keV to 43\% at 10~MeV.

To understand the influence of unit thickness, it is important 
to understand that secondary particles do leave the detector and 
therefore the energy of the incident $\gamma-$ray is not fully 
deposited in the  detector region. There exist two different
classes of secondary particles. First those that 
are detectable by an active shield surrounding the detector. Usually a shield
covers sides and bottom, but not the surface 
through which photons enter the detector because such a shield would interact with 
incoming photons. The direction in which the incoming photons stream we will call 
the "forward direction". Secondary particles leaving the 
detector in the direction of the shield can excluded from the analyses, when the shield is able to detect them
or they can be included in the analyses if the shield gives sufficient energy information. 
Thus in principle these events do not reduce energy resolution, but, they reduce the efficiency. 
The second class are secondary particles which leave against the stream of incoming 
photons. We call this the "upward direction". Since there is no way to detect these 
photons these events are lost for analyses.
The drop in the efficiency at higher incident energy is mainly due to the 
fact that the energy of the incident $\gamma-$ray is not fully deposited 
in the detector region. Therefore the reconstructed energy is lower than 
the incident energy. Figure \ref{PercOutSide} shows the percentage
of events leaving the detector in the upward direction. These are mainly caused 
by pair production events. The right side shows the 
mean energy lost in this way.
 At 662~keV  $\sim$5\% of the events loose on average 
 217~keV. At higher energies the importance
of  this effect increases. At 10~MeV $\sim$23\% of events 
loose on average 700~keV. As already mentioned, there is no remedy for 
this and it is the main cause of the efficiency drop in calorimeters at high energies.
For 0.5~cm thick detectors the energy resolution degrades at higher 
energies. The increase of multiple interactions together with the steep weighting 
potential gradient close to the anodes which 
affects about 50\% of the volume of 0.5~cm thick detectors reduces the energy resolution significantly.

\begin{figure}
\centering
\includegraphics[width=6.0in]{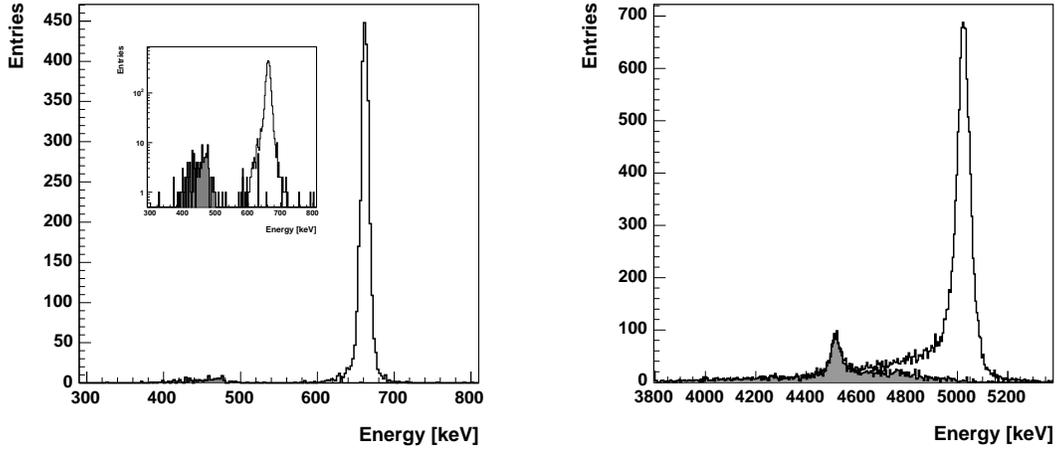}
\caption{Energy
spectra for 662~keV (left side) and 5~MeV (right side) incident energy 
of a full calorimeter with single detector thickness of 1.5~cm. The shaded 
area gives the reconstructed energy of events where 
secondary particles leave the detector in the upward direction, from 
which the primary photons came.
The low energy peak in the spectra is caused by
pair production, when one electron or positron leaves the 
detector in the upward direction. The small panel shows the 662 keV spectra in logarithmic y-axis. }
\label{Spectra5MeVFullCalorimeter}
\end{figure}

In Figure \ref{Spectra5MeVFullCalorimeter}, an energy
spectrum is shown for a full calorimeter with single detector units 
of 1.5~cm thickness for 5~MeV incident energy. The shaded 
area marks events where 
secondary particles leave the detector in upward direction.  

\begin{figure}
\centering
\includegraphics[width=3.0in]{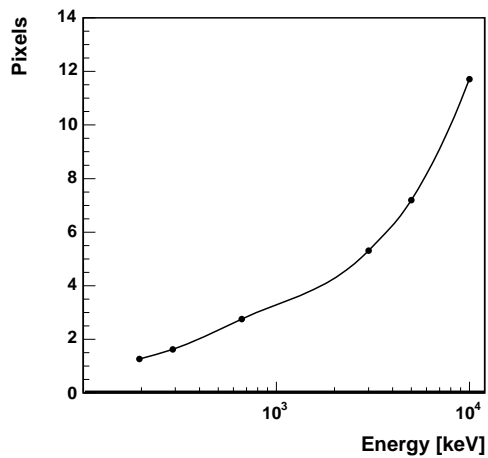}
\caption{Mean number of pixels with measured signal above trigger threshold
as a function of incident energy for a 1.5~cm thick detector.}
\label{NumPixel}
\end{figure}

With increasing energy the 
number of interactions increases and the  readout noise increases 
proportional to  
$\sqrt{n_{pixel}} \times noise$ where $n_{pixel}$ is the number 
of triggered pixels. In Figure \ref{NumPixel} the mean 
number of triggered pixels per incident gamma is shown as a function
of incident energy for a 1.5~cm thick detector. 
From $\sim$200~keV to 10~MeV the mean number of pixels 
increases from around 1 to about 12. For 0.5~cm and 1.0~cm
thick detectors, these values change by less than 2.5\%.

\begin{figure}
\centering
\includegraphics[width=3.0in]{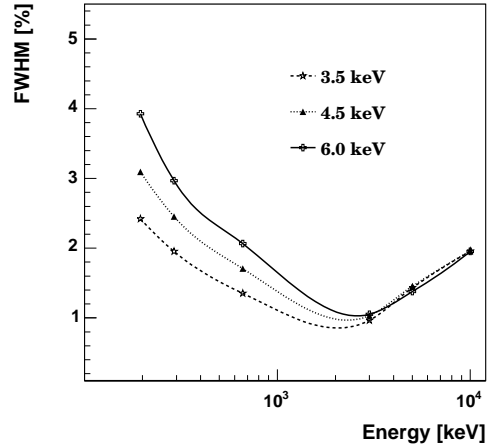}
\caption{Energy resolutions for full calorimeters with 1.5~cm thick single detectors 
as a function of incident energy with 3.5~keV, 4.5~keV and 6~keV FWHM electronic noise  added.}
\label{Noise}
\end{figure}

Finally, Figure \ref{Noise} shows the energy resolution of a full calorimeter
(1.5~cm thick detector units) for single channel noise levels of 
3.5~keV, 4.5~keV and 6~keV. Above 2~MeV  
the contribution of single-channel noise is not anymore  significant while at lower energies the
the overall performance can be improved by reducing 
single-channel noise.
Summarizing it can be said that the efficiency of a full calorimeter is mainly limited by secondary particle leaving the detector in the upward direction while energy resolution is limited by the steep weighting potential near the anode pixels and the signal loss caused by  the trigger threshold.

\section{Conclusion}
\label{con}
In our simulations we have considered weighting potential, electron and hole 
trapping. The results of a single detector unit are in good agreement
with experimental data. Therefore, precipitates and crystal inhomogeneities 
play a major role only for underperforming detector regions. 

The studies in this paper show, that CZT 
 has to be seriously considered as 
a detector material  for the design of 
future calorimeters which operate in the range from several~keV up to several MeV.
We have  simulated three different calorimeter designs  
varying the thickness of the individual detector and the number of layers,
keeping constant the size of the full calorimeter constant 
(10.33~$\times$~10.33~$\times$~45~cm$^3$).

Calorimeters built out of 1.0~cm and 1.5~cm thick units exhibit almost 
identical performance with energy resolutions of about 1.7\% at 662~keV and 
about 1.9\% at 10 MeV.
The energy resolution is mainly limited by the energy loss 
owing to secondary particles leaving the detector upward against the direction 
from which the incident radiation enters the detector and by  the effects of the steep change of the weighting
potential close to the anode.   Going to higher energies the resolution is 
also negatively influenced by the increasing number of events which  have several 
interactions below one pixel because in these cases DOI corrections are not well defined. 

The principal result of our study is the fact that a 0.25~cm thick region below the anodes
of the detector units exhibits a substantially poorer energy resolution
than the rest of the detector volume.  Reducing the portion of this low resolution volume increases the overall performance. A good calorimeter 
requires that the single detector unit is much thicker than the pixel pitch
and of course much thicker than passive material. (mounting, cables and front-end electronics).

Therefore, calorimeters built with 1.0~cm or 1.5~cm thick CZT crystals have significantly better resolution than those with 0.5~cm thin detector units.

The energy resolution of a CZT calorimeter is shown in Figure \ref{EffizienzFWHM}. The resolution is limited  a) 
by the steep weighting potential gradient close 
to the anodes, 
which dominates in thin detectors and impairs the 
 correction of the depth of interaction (DOI), b) by the fact, that the incident energy is not totally deposited  
within the detector area because secondary particles leave the detector upward against the direction 
from which the incident radiation enters the detector c) because signals 
induced in neighboring pixels when the interaction is near the pixel edges 
and near the anodes with intensity below the trigger threshold are lost 
for signal reconstruction leading to systematically reduced signals.

We found that the energy resolution of a CZT calorimeter will lie  
 between that of CsI-detectors and Ge detectors.
In terms of spatial resolution solid state Ge and CZT detectors are 
of course much better than scintillation detectors. 
Where compactness 
and weight are issues as in a next generation space-born $\gamma$-ray telescope, Cadmium Zinc Telluride (CZT) detectors
might be the detector option of choice.


\begin{thebibliography}{00}
\bibitem{He2005}
F.~Zhang, Z.~He, G.~Knoll, D.~Wehe, J.~Berry, \emph{3D Position Sensitive CdZnTe Spectrometer Performance Using Third Generation VAS/TAT Readout
Electronics}, IEEE Transactions on Nuclear Science, Vol. 52, No. 5, October 2005
\bibitem{Matt}
J. Matteson, \emph{CZT detectors with 3-D readout for gamma-ray spectroscopy and imaging},  Proc. SPIE Vol. 4784 (2002)
\bibitem{Macri}
J. R. Macri, B. Donmez, M. Widholm, L.-A. Hamel, M. Julien, T. Narita, J. M. Ryan and M. L. McConnell, \emph{Single-Sided CZT Strip Detectors}  Proc. SPIE, 5501, 208 (2004). 
\bibitem{DOE}
B. Doenmez, J. R. Macri, M. L. McConnell, J. M. Ryan, M. Widholm, T. Narita, and L.-A. Hamel, Proc. SPIE, 5922, 62 (2005). 
\bibitem{G4}
S.~Agostinelli et al., \emph{Geant4--a simulation toolkit}, NIM A506, 250-303,2003
\bibitem{MEGAProject}
G.~Kanbach et al., \emph{The MEGA project}, New Astronomie Reviews 48, 275-280, 2004.
\bibitem{Knoe}
J.~Knoedleseder, P.~Jean, V.~Lonjou et al. 2005, A\&A, in press
\bibitem{TIGERProject}
T.J.~O'Neill et al., \emph{Development of the TIGRE Compton Telescope for Intermediate-Energy Gamma-Ray Astronomy}, IEEE Trans. Nucl. Sci. 50 2 , 251, 1999.
\bibitem{MEGACalib}
R.~Andritschke et al., \emph{The calibration setup of the MEGA prototype at the high intensity c-ray source}, New Astronomie Reviews 48, 281-285, 2004.
\bibitem{MEGACalorimeter}
F.~Schopper et al., \emph{CsI calorimeter with 3-D position resolution}, NIM A442, 394-399, 2000.
\bibitem{Gretina}
K.~Vetter, \emph{Performance of the GRETA prototype detectors}, NIM A452, 105-114, 2000.

\bibitem{Gretina1}
K.~Vetter, \emph{Three-dimensional position sensitivity in two-dimensionally segmented HP-Ge detectors}, NIM A452, 223-238, 2000.
\bibitem{evProduct}{eV Products, 373 Saxonburg Boulevard, Saxonburg, PA 16056}
\bibitem{Scho}
W. Shockley, \emph{Currents to conductors induced by a moving point charge}, J. Appl. Phys. 9 (1938) 635
I.R.E., September 1939, 584
\bibitem{Ramo}
S. Ramo, \emph{Current induced by electron motion}, Proceedings of  the 
I.R.E., September 1939, 584
\end{thebibliography}
\end{document}